\documentclass[prb,aps,twocolumn]{revtex4-1}
\usepackage{bm}
\usepackage{amsmath}
\usepackage{amssymb} 
\usepackage{epsfig}
\usepackage{graphicx}
\usepackage{color}
\usepackage{xcolor}
\usepackage{ulem} 
\usepackage{cancel} 
\usepackage{hyperref} 
\renewcommand{\phi}{\varphi}
\renewcommand{\kappa}{\varkappa}
\newcommand{\ket}[1]{\left| #1 \right\rangle}
\renewcommand{\i}{\mathrm i}

\begin{document}

\title{Splitting of Dirac cones in HgTe quantum wells: Effects of crystallographic orientation, interface-, bulk-, and structure-inversion asymmetry}

\author{M.\,V.\,Durnev}
\author{G.\,V.\,Budkin}
\author{S.\,A.\,Tarasenko}

\affiliation{Ioffe Institute, 194021 St.\,Petersburg, Russia}

\begin{abstract}

We develop a microscopic theory of the fine structure of Dirac states in $(0lh)$-grown HgTe/CdHgTe quantum wells (QWs), where $l$ and $h$ are the Miller indices. It is shown that bulk, interface, and structure inversion asymmetry causes the anticrossing of levels even at zero in-plane wave vector and lifts the Dirac state degeneracy. In the QWs of critical thickness, the two-fold degenerate Dirac cone gets split into non-degenerate Weyl cones. The splitting and the Weyl point positions dramatically depend on the QW crystallographic orientation. We calculate the splitting parameters related to bulk, interface, and structure inversion asymmetry and derive the effective Hamiltonian of the Dirac states. Further, 
we obtain an analytical expression for the energy spectrum and discuss the spectrum for (001)-, (013)- and (011)-grown QWs.
 
\end{abstract}
 
\maketitle

\section{Introduction}

Heterostructures containing band-inverted compound HgTe are in the focus of modern research in solid state physics. 
Depending on heterostructure design, particularly the thickness of HgTe layer, they host a variety of phases 
including the phases of three-dimensional and two-dimensional (2D) topological insulators, 2D gapless semiconductor, 
2D semimetal, etc~\cite{Qi2011,Kvon2020}. Of special interest is the 2D gapless phase with linearly-dispersion Dirac fermions that is realized 
in HgTe quantum wells (QWs) of critical thickness, i.e., at the point of trivial insulator -- topological insulator transition~\cite{Buttner2011,Tarasenko2015}.
 
In the model of centrosymmetric heterostructure, the Dirac cones in HgTe/CdHgTe QWs are 2-fold degenerate yielding
the 4-fold degenerate Dirac point at $\bm k = 0$ in the QW of critical thickness~\cite{Buttner2011}.   
Here, $\bm k$ is the in-plane wave vector.
Bulk inversion asymmetry (BIA) related to the lack of an inversion center in host zinc-blend crystal, 
interface inversion asymmetry (IIA) related to anisotropy of chemical bonds at interfaces, and structure inversion asymmetry (SIA)
lift the Dirac state degeneracy~\cite{Dai2008,Konig2008,Winkler2012,Weithofer2013,Tarasenko2015,Orlita2011,Zholudev2012,Olbrich2013,Minkov2016,Durnev2016}. 
This splitting is contributed by canonical $\bm k$-linear Rashba~\cite{Rashba1960,Vasko1979,Bychkov1984} 
and Dresselhaus~\cite{Dresselhaus1955,Dyakonov1986,Pikus1988,Rashba1988} terms 
as well as the term lifting the 4-fold degeneracy at $\bm k =0$~\cite{Dai2008,Konig2008,Winkler2012,Weithofer2013,Tarasenko2015}. 
The anticrossing gap at $\bm k = 0$ was found to be quite large in (001)-grown QWs and originates mostly from light-hole--heavy-hole 
mixing at the QW interfaces~\cite{Tarasenko2015,Minkov2016}.  
 
Many experiments, however, are being carried out on HgTe/CdHgTe structures grown along low-symmetry crystallographic directions, 
such as [013] and [012], see Refs.~\onlinecite{Zholudev2012,Olbrich2013,Minkov2016,Dantscher2015,Dantscher2017,Minkov2017}. 
The choice of crystallographic orientations is dictated by technology: MBE growth of HgTe and CdHgTe layers 
on low-symmetry (lattice-mismatch) GaAs surface enables one to obtain high-quality structures~\cite{Dvoretsky2020}. 
This motivates theoretical studies of low-symmetry QWs~\cite{Minkov2017,Raichev2012,Budkin2022}. 

Here, we develop a microscopic theory of the fine structure of Dirac states in HgTe/CdHgTe QWs taking account IIA, BIA, and SIA coupling.
We show that the energy spectrum in the QW of the critical thickness dramatically depends on the QW crystallographic orientation and 
calculate the splitting parameters. We explore the class of $(0lh)$-grown QWs, where $l$ and $h$ are the Miller indices, and study 
how the fine structure evolves from (001)- to (013)-, and (011)-grown QWs.

\section{Fine structure of Dirac states}

The Dirac states in HgTe/CdHgTe QWs of critical and close-to-critical thickness are formed from the electron-like $\ket{ E1, \pm 1/2}$ and 
heavy-hole $\ket{H1, \pm 3/2}$ states~\cite{Bernevig2006,Gerchikov1989}. The corresponding basis functions at $\bm k = 0$ have the form
\begin{align}\label{E1H1}
& \ket{ E1, + 1/2} = f_1(z) \ket{\Gamma_6, + 1/2} + f_4(z) \ket{\Gamma_8, + 1/2} \,, \nonumber \\
& \ket{ H1, + 3/2} = f_3(z) \ket{\Gamma_8, + 3/2} \,, \nonumber \\
& \ket{ E1, - 1/2} = f_1(z) \ket{\Gamma_6, - 1/2} + f_4(z) \ket{\Gamma_8, - 1/2} \,, \nonumber \\
& \ket{ H1, - 3/2} = f_3(z) \ket{\Gamma_8, - 3/2} \,, 
\end{align} 
where $\bm k = (k_x, k_y)$ is the in-plane wave vector, $f_j (z)$ ($j=1,3,4$) are the envelope functions, $z$ is the growth direction, 
$\ket{\Gamma_6, m}$ ($m = \pm 1/2$) and $\ket{\Gamma_8, m}$ ($m = \pm 1/2 , \pm 3/2$) are the Bloch amplitudes of the $\Gamma_6$ 
and $\Gamma_8$ bands, respectively, in the Brillouin zone center. We consider $(0lh)$-oriented QWs and use the QW-related coordinate frame
 $x \parallel [100]$, $y \parallel [0h\bar{l}]$, and $z \parallel [0lh]$. 
 
The effective $4 \times 4$ Hamiltonian, which describes the coupling of the basis states~\eqref{E1H1} and formation of the Dirac-like spectrum,
can be derived in the $\bm k$$\cdot$$\bm p$ theory, see Sec.~\ref{Sec_6band}.  
Taking into account bulk, structure, and interface inversion asymmetry in $(0lh)$-grown QWs, one can present the effective Hamiltonian in the form
\begin{equation}
\label{Hqw}
H = H_0 + H_{\rm IIA/BIA} + H_{\rm SIA} \,,
\end{equation}
where
\begin{equation}
\label{BHZ}
H_0 = \left(
\begin{array}{cccc}
\delta & \i A k_+ & 0 & 0 \\
- \i A k_- & -\delta &  & 0 \\
0 & 0 & \delta  & - \i A k_- \\
0 & 0 & \i A k_+ & - \delta
\end{array}
\right)
\end{equation} 
is the $\bm k$-linear Bernevig-Hughes-Zhang Hamiltonian (2D Dirac Hamiltonian) ~\cite{Bernevig2006}, $A$ is a parameter determining the velocity of Dirac fermions, $\delta$ is the energy distance between the $E1$ and $H1$ subbands in the absence of mixing. 

Interface inversion asymmetry related to anisotropy of chemical bonds at the interfaces and bulk inversion asymmetry  
related to the lack of inversion center in host crystal lead to a mixing of the basis states.  
This mixing at $\bm k = 0$ is described by the Hamiltonian
\begin{equation}\label{HIIABIA} 
H_{\rm IIA/BIA} = \left(
\begin{array}{cccc}
0 &  - \eta \sin 2\theta & 0 & \i \gamma \cos 2\theta \\
 - \eta \sin 2\theta  & 0 & \i \gamma \cos 2\theta  & 0 \\
0 & - \i \gamma \cos 2\theta & 0 &  \eta \sin 2\theta \\
- \i \gamma \cos 2\theta  & 0 &  \eta \sin 2\theta & 0 
\end{array}    
\right) , 
\end{equation}
where $\eta$ and $\gamma$ are mixing parameters, $\theta = {\rm arctan} (l/h)$ is the angle between the QW growth direction 
$[0lh]$ and the $[001]$ axis. The angles $\theta = 0$, ${\rm arctan} (1/3) \approx 0.321$, and $\pi /4$ correspond to $[001]$, 
$[013]$, and $[011]$ growth directions, respectively.

Structure inversion asymmetry in $(0lh)$-oriented QWs grown from cubic materials also mixes the basis states at $\bm k = 0$,
which is described by the Hamiltonian
\begin{equation}\label{HSIA}
H_{\rm SIA} = \left(
\begin{array}{cccc}
0 & \i \chi \sin 4 \theta & 0 & \zeta \sin^2 2 \theta \\
- \i \chi \sin 4 \theta  & 0 & \zeta \sin^2 2 \theta  & 0 \\
0 & \zeta \sin^2 2 \theta & 0  & \i \chi \sin 4 \theta \\
 \zeta \sin^2 2 \theta & 0 & - \i \chi \sin 4 \theta & 0
\end{array}
\right) \,.
\end{equation}
The mixing parameters $\chi$ and $\zeta$  are nonzero if both structure inversion asymmetry and the cubic shape of lattice unit cells are taken into account. Note also that $H_{\rm SIA}$ vanishes in (001)-grown QWs.

The parameters $\eta$, $\gamma$, $\chi$, and $\zeta$ are calculated in Sec.~\ref{Sec_6band} in the framework of the 6-band $\bm k$$\cdot$$\bm p$ model.  An estimation for HgTe/Cd$_{0.7}$Hg$_{0.3}$Te QWs with the critical thickness ${d_c \approx 6.7}$~nm gives
 $\eta,\gamma \sim 5$~meV and $\zeta,\chi \sim 0.1$~meV in the electric field $E_z=15$~kV/cm. 

The Hamiltonians $H_{\rm IIA/BIA}$~\eqref{HIIABIA} and $H_{\rm SIA}$~\eqref{HSIA} depend on the QW growth direction defined by the $\theta$ angle. 
Straightforward diagonalization of the Hamiltonian~\eqref{Hqw} yields four dispersion branches
\begin{eqnarray}
\label{cones_total}
E_{1,4} &=& \mp \sqrt{ \delta^2 + \gamma_\theta^2 + \eta_\theta^2 + \zeta_\theta^2 + \chi_\theta^2 + A^2 k^2 + 2A K } \:, \nonumber \\
E_{2,3} &=& \mp \sqrt{ \delta^2 + \gamma_\theta^2 + \eta_\theta^2 + \zeta_\theta^2 + \chi_\theta^2 + A^2 k^2 - 2A K } \:,
\end{eqnarray}
where $k^2 = k_x^2 + k_y^2$,
\begin{equation}
K = \sqrt{(\gamma_\theta^2 + \zeta_\theta^2) k^2 + \left(\chi_\theta k_x + \eta_\theta k_y\right)^2}\:,
\end{equation}
and
\begin{eqnarray}
\gamma_\theta &=& \gamma \cos 2\theta\:,~~~\eta_\theta = \eta \sin 2\theta\:, \nonumber \\
\zeta_\theta &=& \zeta \sin^2 2\theta\:,~~\chi_\theta = \chi \sin 4 \theta\:. 
\end{eqnarray}
In the following subsections we analyze the fine structure of Dirac states in QWs with different crystallographic orientations for different mixing mechanisms.

\subsection{Interface and bulk inversion asymmetry}

Dirac states in HgTe/CdHgTe QWs with symmetric confinement potential are described by the Hamiltonian $H = H_0 + H_{\rm IIA/BIA}$ with
$H_0$ and $H_{\rm IIA/BIA}$ given by Eqs.~\eqref{BHZ} and~\eqref{HIIABIA}, respectively.
In this case, Eq.~\eqref{cones_total} yields
\begin{eqnarray}
\label{cones_BIAIIA}
E_{1,4} &=& \mp \sqrt{\delta^2 + \gamma_\theta^2 + \eta_\theta^2 + A^2 k^2 + 2A \sqrt{\gamma_\theta^2 k^2 +  \eta_\theta^2 k_y^2}}\:, \nonumber \\
E_{2,3} &=& \mp \sqrt{\delta^2 + \gamma_\theta^2 + \eta_\theta^2 + A^2 k^2 - 2A \sqrt{\gamma_\theta^2 k^2 +  \eta_\theta^2 k_y^2}}\:. \nonumber \\
\end{eqnarray}

\begin{figure}[htpb]
\includegraphics[width=0.47\textwidth]{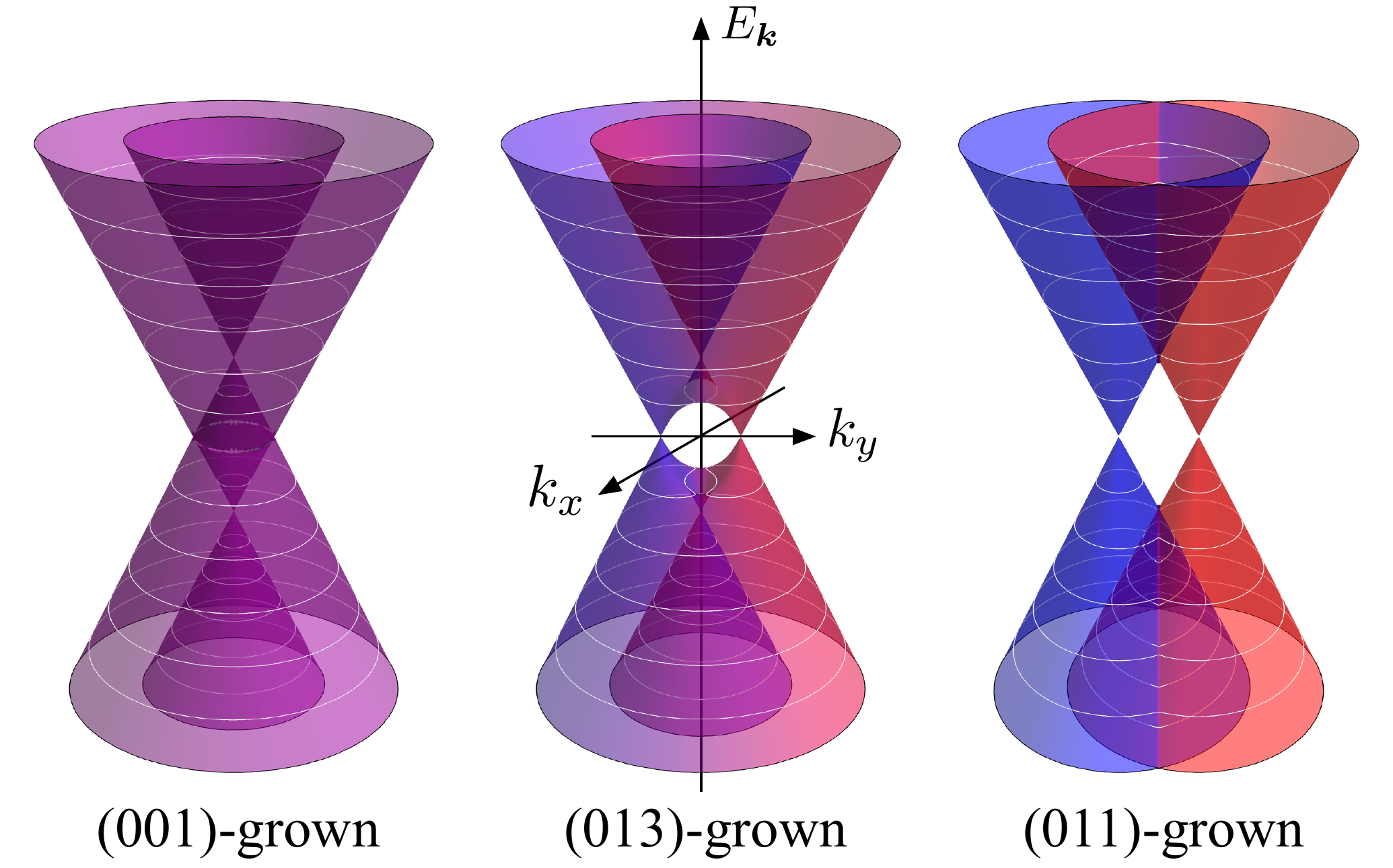}
\caption{\label{fig1} 
Energy spectra of Dirac states in (001)-, (013)- and (011)-grown HgTe/CdHgTe QWs of critical thickness with interface- and bulk-inversion asymmetry included. The spectra are calculated after Eq.~\eqref{cones_BIAIIA} for $\delta = 0$ and $\gamma = \eta$. Color decodes the peudospin projection onto the QW normal (see text for details): blue and red correspond to $\sigma_z = -1$ and $\sigma_z = +1$, respectively, whereas purple corresponds to $\sigma_z = 0$. 
}
\end{figure}

\begin{figure}[htpb]
\includegraphics[width=0.47\textwidth]{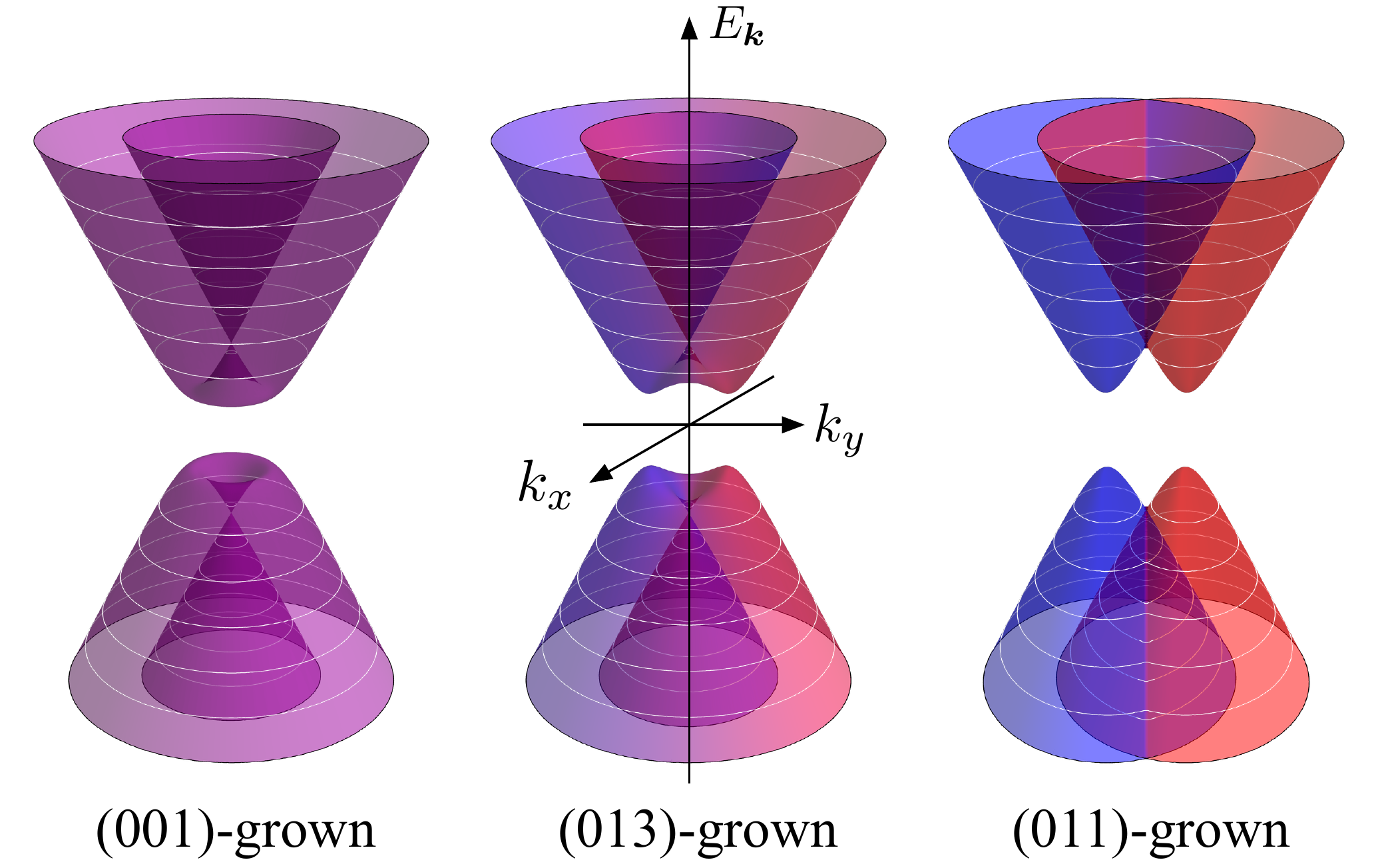}
\caption{\label{fig2} 
Energy spectra of Dirac states in (001)-, (013)- and (011)-grown HgTe/CdHgTe QWs of close-to-critical thickness with interface- and bulk-inversion asymmetry included. The spectra are calculated after Eq.~\eqref{cones_BIAIIA} for $\gamma = \eta = 2 \delta$. Color decodes the peudospin projection onto the QW normal: blue and red correspond to $\sigma_z = -1$ and $\sigma_z = +1$, respectively, whereas purple corresponds to $\sigma_z = 0$.
}
\end{figure}
 
Figure~\ref{fig1} shows the energy spectra of Dirac states in (001), (013), and (011) QWs of critical thickness ($\delta = 0$) 
with the IIA/BIA term included. The (001) orientation corresponds to $\theta = 0$.
In this case, the energy spectrum consists of two non-degenerate (Weyl) cones shifted vertically (along the energy axis) with respect to each other~\cite{Tarasenko2015}. The Weyl points are located at $\bm k =0$ and the energies $E = \pm \gamma$. 
The (011) orientation corresponds to $\theta = \pi/4$. In such QWs, the IIA/BIA interaction splits the Dirac cone into two Weyl cones shifted along $k_y$
with respect to each other. The Weyl points are located at $\bm k = (0, \pm \eta/A)$.  The spectrum in (013) and general $(0lh)$ QWs 
is an intermediate case between the spectra in (001) and (011) structures. Now, there are four Weyl points in the energy spectrum. 
Two points are located at $\bm k = (0, \pm \sqrt{\gamma_\theta^2  + \eta_\theta^2}/A)$ and zero energy, while the other two points are at
$\bm k = 0$ and the energies $E = \pm \sqrt{\gamma_\theta^2 + \eta_\theta^2}$. 

The color in Fig.~\ref{fig1} decodes the projection of pseudospin onto the QW normal $\sigma_z$ defined by 
$\sigma_z = |c_1|^2 + |c_2|^2 - |c_3|^2 - |c_4|^2$, where $c_j$ are the coefficients of decomposition of a wave function $\psi$ over the basis functions~\eqref{E1H1}.  It illustrates the relative contribution of the ``spin-up'' ($\ket{E1,+1/2}$ and $\ket{H1, +3/2}$) and ``spin-down'' ($\ket{E1,-1/2}$ and $\ket{H1, -3/2}$)  blocks in the given state $\psi$. In (001) QWs, the Weyl cones are formed by the ``spin-up'' and ``spin-down'' blocks in equal portions
and $\sigma_z = 0$ (purple color) for all eigen states. In contrast, the split Weyl cones in (011) QWs are formed by pure  ``spin-up'' and ``spin-down'' states and characterized by $\sigma_z = +1$ (red color) and $\sigma_z = -1$ (red color) pseudospin projections.

The energy spectra of (001), (013), and (011) QWs of close-to-critical thickness (with the gap $2 | \delta |$) are shown in Fig.~\ref{fig2}. 
In (001) QWs, the spectrum is given by $E = \pm \sqrt{\delta^2 + (A |\bm k| \pm \gamma)^2}$ and the band extrema are 
situated at the loop with $|\bm k| = |\gamma/A|$. In (011) QWs, the spectrum has the form $E = \pm \sqrt{\delta^2 + A^2k_x^2 + (Ak_y \pm \eta)^2}$
and consists of the branches with the $\sigma_z = \pm 1$ pseudospin projections. In general case of $(0lh)$ orientation, e.g., (013), the band extrema are
situated at the points $\bm k = (0, \pm \sqrt{\gamma_\theta^2 + \eta_\theta^2}/A)$. The contours of constant energy are toric sections, in particular, at $|E| > \sqrt{\delta^2 + \gamma_\theta^2 + \eta_\theta^2}$ the isoenergy contours are ovals elongated along $k_y$.

\subsection{Structure inversion asymmetry}

Here, we study the influence of structure inversion asymmetry on the energy spectrum of Dirac states. For this purpose, we consider the Hamiltonian 
$H = H_0 + H_{\rm SIA}$ with $H_0$ and $H_{\rm SIA}$ given by Eqs.~\eqref{BHZ} and \eqref{HSIA}, respectively. In this case, Eq.~\eqref{cones_total} yields
\begin{eqnarray}
\label{cones_SIA}
E_{1,4} &=& \mp \sqrt{\delta^2 + \zeta_\theta^2 + \chi_\theta^2 + A^2 k^2 + 2A \sqrt{\zeta_\theta^2 k^2 +  \chi_\theta^2 k_x^2}}\:, \nonumber \\
E_{2,3} &=& \mp \sqrt{\delta^2 + \zeta_\theta^2 + \chi_\theta^2 + A^2 k^2 - 2A \sqrt{\zeta_\theta^2 k^2 +  \chi_\theta^2 k_x^2}}\:. \nonumber \\
\end{eqnarray}

\begin{figure}[htpb]
\includegraphics[width=0.47\textwidth]{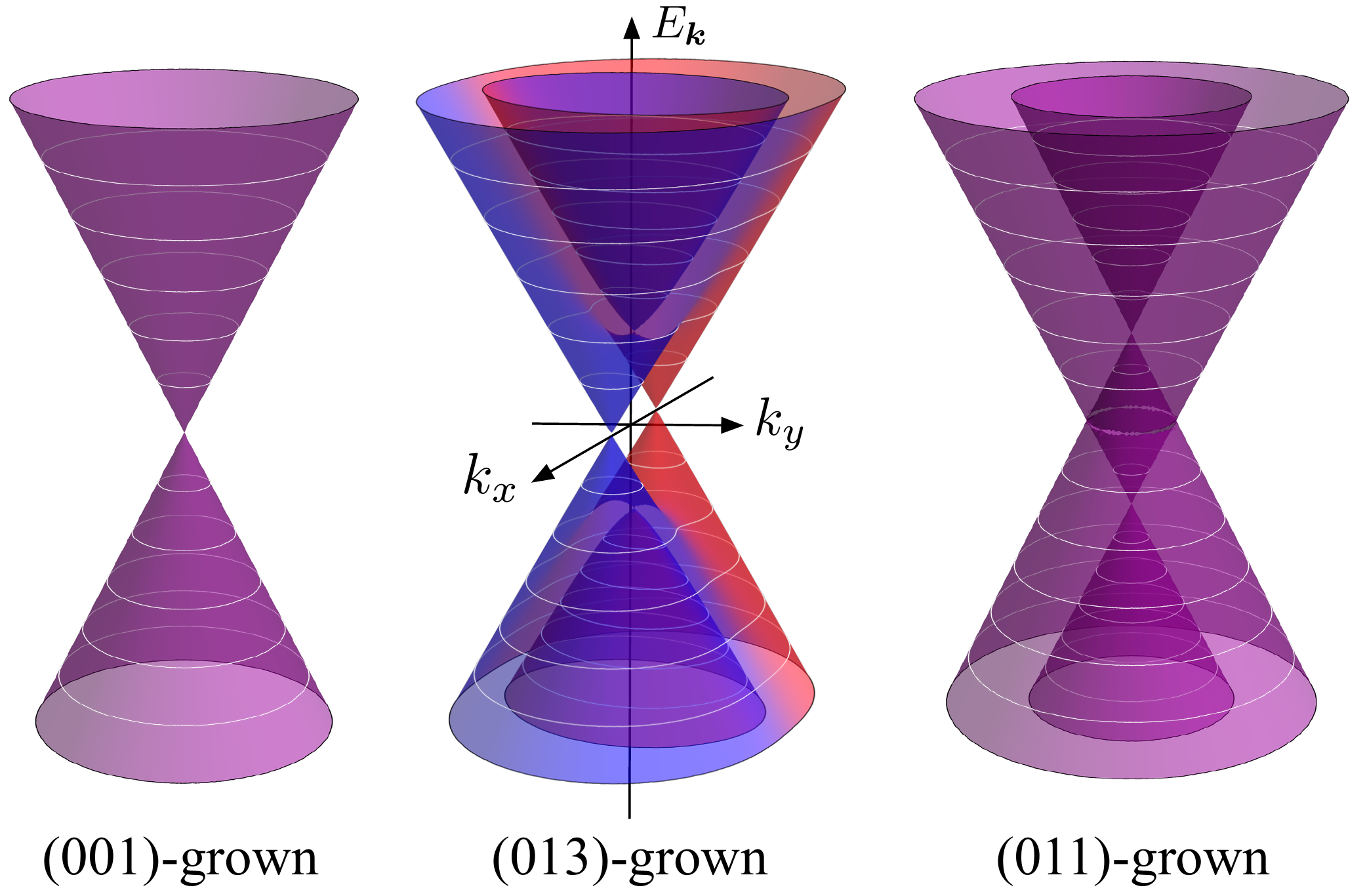}
\caption{\label{fig3} 
Energy spectra of Dirac states in (001)-, (013)- and (011)-oriented QWs of critical thickness with structure-inversion asymmetry included.
The spectra are calculated after Eq.~\eqref{cones_SIA} for $\delta = 0$ and $\zeta = \chi$. Color decodes the peudospin projection onto the QW normal: blue and red colors correspond to $\sigma_z = -1$ and $\sigma_z = +1$, respectively, whereas purple color corresponds to $\sigma_z = 0$. 
}
\end{figure}

Figure~\ref{fig3} shows the energy dispersions given by Eq.~\eqref{cones_SIA} for (001), (013), and (011) QWs of critical thickness ($\delta = 0$). 
In (001) QWs, structure inversion asymmetry does not contribute to the mixing of the basis states at $\bm k =0$. As a result, the point $\bm k = 0$
remains four-fold degenerate just as it is in the Bernevig-Hughes-Zhang model.
In QWs of other orientations, the SIA interaction lifts the four-fold degeneracy at $\bm k =0$ and splits the Dirac cone. Generally,
there are four Weyl points in the energy spectrum located at $\bm k = (\pm \sqrt{\zeta_\theta^2 + \chi_\theta^2 }/A, 0)$ and zero energy and
at $\bm k = 0$ and the energies $E = \pm \sqrt{\zeta_\theta^2 + \chi_\theta^2 }$, respectively. 
Interestingly, the SIA interaction in (011) QWs splits the Dirac cone in a way similar to the IIA/BIA interaction does in (001) QWs.

In QW structures with a gap (not shown),  the band extrema are located at $\bm k = (\pm \sqrt{\zeta_\theta^2 + \chi_\theta^2 }/A, 0)$, in particular, 
at $\bm k = 0$ in (001) QWs and at the loop with $|\bm k| = | \zeta/A |$ in (011) QWs.

\subsection{Interplay of IIA/BIA and SIA}

\begin{figure}[htpb]
\includegraphics[width=0.47\textwidth]{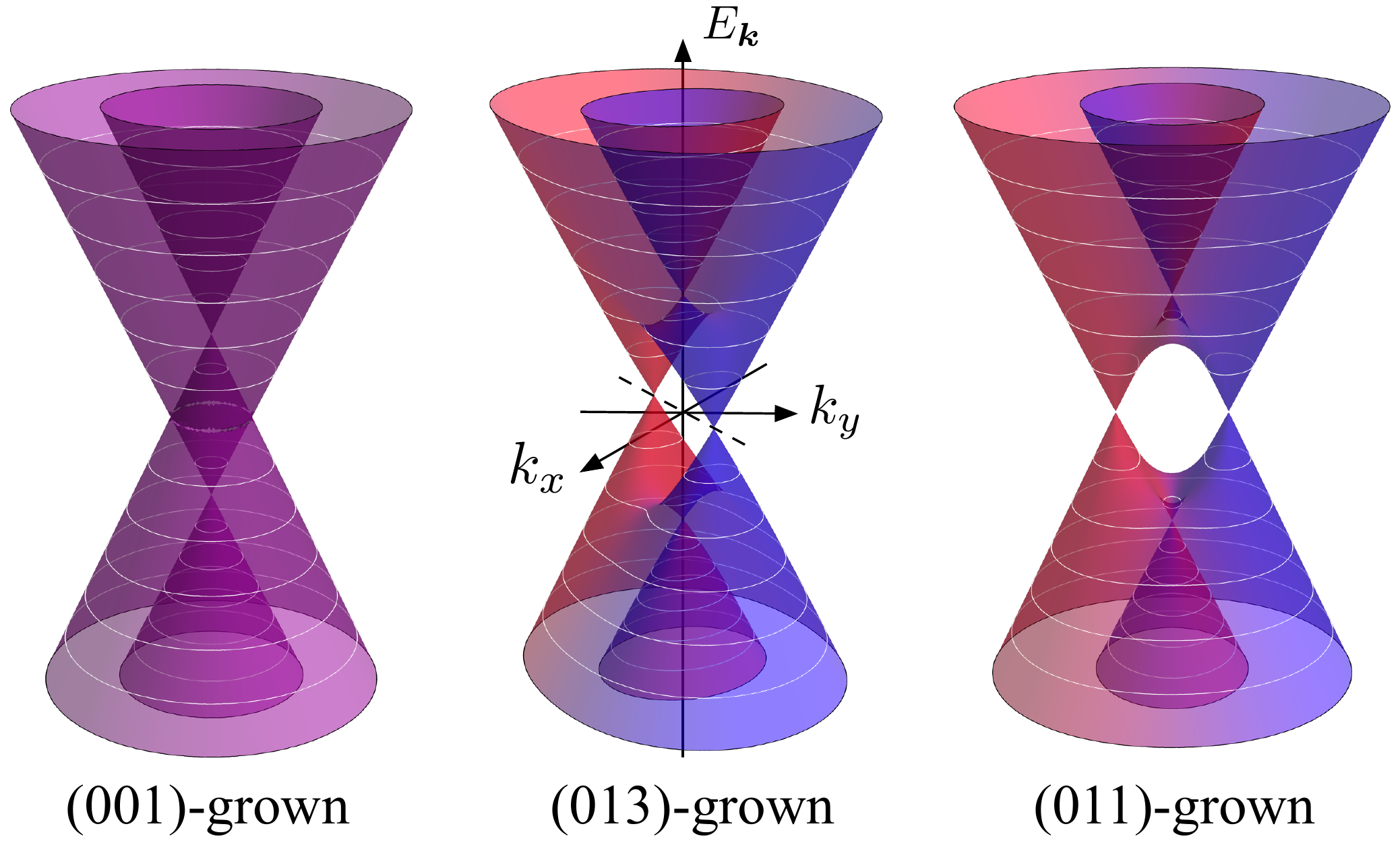}
\caption{\label{fig4} 
Energy spectra of Dirac states in (001)-, (013)- and (011)-grown HgTe/CdHgTe QWs of critical thickness with interface-, bulk-, and structure-inversion asymmetry included. The spectra are calculated after Eq.~\eqref{cones_total} for $\delta = 0$ and $\gamma = \eta = \zeta = \chi$. Color decodes the peudospin projection onto the QW normal: blue and red correspond to $\sigma_z = -1$ and $\sigma_z = +1$, respectively, whereas 
purple corresponds to $\sigma_z = 0$. 
}
\end{figure}

In real QW structures, all types of asymmetry, including bulk, interface, and structure inversion asymmetry, are present. 
The dispersion branches in that case are given by the general Eq.~\eqref{cones_total}. 
Figure~\ref{fig4} shows the energy spectra of Dirac states in such QWs with asymmetric confinement potential and grown along different 
crystallographic orientations.

In (001) QWs, the splitting of the Dirac cone at $\bm k = 0$ is determined by the BIA/IIA term and the energy spectrum coincides with the one shown in Fig.~\ref{fig1}. The spectrum of asymmetric (011) QWs is qualitatively similar to the spectrum of symmetric $(013)$ QWs. There are four Weyl points:
two of them located at $\bm k = (0, \pm \sqrt{\zeta_\theta^2  + \eta_\theta^2}/A)$ and zero energy and the other two located at $\bm k = 0$ 
and the energies $E = \pm \sqrt{\zeta_\theta^2  + \eta_\theta^2}$. 

The spectrum of a general $(0lh)$-grown QW with asymmetric confinement potential is shown in the central panel in Fig.~\ref{fig4}.
The Weyl points at zero energy are located at the wave vectors 
\begin{equation}
\bm k = \pm \sqrt{\frac{\gamma_\theta^2 + \eta_\theta^2 + \zeta_\theta^2 + \chi_\theta^2}{A^2 \left(\eta_\theta^2 + \chi_\theta^2 \right)}} \left(\chi_\theta , \eta_\theta \right)\:.
\end{equation}
Interestingly, the position of these points in the $\bm k$ space is not pinned to a specific in-plane direction. The angle between the line connecting the Weyl points and the $k_x$ axis, $\arctan (\eta_\theta/\chi_\theta)$, depends on the SIA parameter $\chi$ and, therefore, can be controlled by an external electric field applied along the QW normal, e.g., by gate voltage.

\section{6-band \textbf{kp} theory}\label{Sec_6band}

In this section, we calculate the energy spectrum and derive the parameters of the effective Hamiltonian~\eqref{Hqw} 
using the extended 6-band $\bm k$$\cdot$$\bm p$ theory.
The conduction-band and valence-band states in HgTe/CdHgTe structure are mainly formed from the $\Gamma_6$ and $\Gamma_8$ bands, which are 2-fold and 4-fold degenerate, respectively, 
at $\bm k = 0$ in the bulk crystal~\cite{Novik2005, Gerchikov1989, Bernevig2006}. 
Taking into account $\bm k$$\cdot$$\bm p$ mixing, deformation interaction, and interface mixing, we present the corresponding 6-band Hamiltonian 
in the form
\begin{equation}\label{H6band}
\mathcal H = \mathcal H_{\rm kp} + \mathcal H_{\rm def} + \mathcal H_{\rm int}  =\left(
\begin{array}{c c}
\mathcal H_{66} & \mathcal H_{68} \\
\mathcal H^\dag_{68} & \mathcal H_{88}
\end{array}
\right) ,
\end{equation}
where $\mathcal H_{66}$ is the 2$\times$2 matrix in the $\ket{\Gamma_6, +1/2}$, $\ket{\Gamma_6, - 1/2}$ basis, 
$\mathcal H_{88}$ is the 4$\times$4 matrix in the $\ket{\Gamma_8, +3/2}$, $\ket{\Gamma_8, + 1/2}, \ket{\Gamma_8, - 1/2}, \ket{\Gamma_8, - 3/2}$ 
basis, $\mathcal H_{68}$ is the 2$\times$4 matrix, which couples the $\Gamma_6$ and $\Gamma_8$ blocks, and $\mathcal H^\dag_{68}$ is
the Hermitian conjugate matrix. 

The isotropic version of the Hamiltonian~$\mathcal H_{\rm kp}$ is the well-known 6-band Kane Hamiltonian which is often used to model the
conduction and valence bands in narrow gap III-V semiconductors. This model, however, is simplified and does not take into account the real 
symmetry ($T_d$ point group) of the zinc-blende lattice. The latter is essential to describe the fine structure of Dirac states. 
Therefore, we go beyond the isotropic Kane Hamiltonian and make use of the so-called extended Kane model~\cite{Winkler_book}, 
which takes into account the cubic shape and inversion asymmetry of the lattice. 

The extended Kane Hamiltonian $\mathcal H_{\rm kp}$ is constructed by the methods of group representation theory~\cite{BirPikus}. 
In the cubic axes $x' \parallel [100]$, $y' \parallel [010]$, and $z' \parallel [001]$, the $\mathcal H_{66}^{\rm kp}$ and 
$\mathcal H_{88}^{\rm kp}$ blocks to the second order in the wave vector $\bm k$ have the form
\begin{equation}\label{H66kp}
\mathcal H_{66}^{\rm kp} = U_6 + \frac{\hbar^2 k^2}{2 m_e'}\:,
\end{equation}
\begin{multline}\label{H88kp}
\mathcal H_{88}^{\rm kp} = U_8 + \frac{\hbar^2}{2m_0} \left[ -\left(\gamma_1' + \frac52 \gamma_2' \right) k^2 + 2 \gamma_2' (\bm J \bm k)^2 +  \right.\\
\left. + 2(\gamma_3' - \gamma_2') \sum_{i \neq j}\{J_i J_j\}_s k_i k_j \right] + \frac{4 \varkappa_0}{\sqrt{3}} \bm V \bm k \:,
\end{multline}
where $U_6$ and $U_8$ are the energies of the $\Gamma_6$ and $\Gamma_8$ bands at $\bm k = 0$,  $\bm k = (k_{x'}, k_{y'}, k_{z'})$ is the wave vector, 
$\gamma_1'$, $\gamma_2'$, $\gamma_3'$, and $m_e'$ are the contributions to the Luttinger parameters and the effective mass, respectively, 
from remote bands and free electron dispersion, $\bm J = (J_ {x'}, J_{y'}, J_{z'})$ is the vector composed of the momentum-3/2 matrices, 
$\bm V = (V_{x'}, V_{y'}, V_{z'})$, where $V_{x'} = \{J_{x'}, J_{y'}^2 - J_{z'}^2\}$ and the other components of $\bm V$ are derived by the cyclic permutation 
of the subscripts, $\{ A, B\}_s = (AB+BA)/2$ is the symmetrized product of the operators $A$ and $B$, and $\varkappa_0$ is a band parameter. 

To construct the $\mathcal H_{68}$ block, one notes that the direct product $\Gamma_6 \times \Gamma_8^*$ is decomposed into the irreducible representations $\Gamma_3 + \Gamma_4 + \Gamma_5$.  The sets $\{k_{x'}, k_{y'}, k_{z'}\}$ and $\{k_{y'} k_{z'}, k_{x'} k_{z'}, k_{x'} k_{y'}\}$ 
transform according to the vector representation $\Gamma_5$ whereas the pair $\{2k_{z'}^2 - k_{x'}^2 - k_{y'}^2, \sqrt{3}(k_{x'}^2 - k_{y'}^2)\}$ 
transforms according to the $\Gamma_3$ representation. The combinations that transform according to the pseudo-vector representation 
$\Gamma_4$ are cubic in $\bm k$ and are not considered. Thus, the $\mathcal H_{68}^{\rm kp}$ block is given by~\cite{Winkler_book}
\begin{equation}\label{H68kp}
\mathcal H_{68}^{{\rm kp} \dag} = \left(
\begin{array}{cc}
\frac{\i }{\sqrt{2}} (B_+ k_+ k_{z'} -P k_-)  &\frac{\i}{3 \sqrt{2}} B_-(2 k_{z'}^2 - k_\parallel^2)  \\
 & \\
\sqrt{\frac23} (\i P k_{z'} + B_+ k_{x'} k_{y'})   & \frac{\i}{\sqrt{6}}(B_+ k_+k_{z'} - Pk_-) \\
- \frac{\i}{\sqrt{6}} B_- (k_{y'}^2 - k_{x'}^2) &  \\
 & \\
  \frac{\i}{\sqrt{6}}(Pk_+ +  B_+ k_- k_{z'} )& 
 \sqrt{\frac23} (\i Pk_{z'} + B_+ k_{x'} k_{y'}) \\
 & +\frac{\i}{\sqrt{6}} B_- (k_{y'}^2 - k_{x'}^2) \\
  & \\
 \frac{\i}{3\sqrt{2}} B_- (k_\parallel^2 - 2 k_{z'}^2) & \frac{\i}{\sqrt{2}}(Pk_+ + B_+k_-k_{z'})
\end{array}
\right) ,
\end{equation}
where $P = \i (\hbar/m_0) p_{cv} $ is the Kane matrix element, $B_{\pm}$ are band parameters, 
$k_\pm = k_{x'} \pm \i k_{y'}$, and $k_\parallel^2 = k_{x'}^2 + k_{y'}^2$. 
Note that the definition of $P$ and $B_{\pm}$ in Eq.~\eqref{H68kp} differs by the factor of $\i$ from that in Ref.~\onlinecite{Winkler_book}.

The extended Kane Hamiltonian given by the blocks~\eqref{H66kp}-\eqref{H68kp} reflects the real symmetry of the zinc-blende lattice
including its cubic shape and the lack of space inversion center. The isotropic centrosymmetric approximation corresponds to 
$\gamma_2' = \gamma_3'$, $\varkappa_0 = 0$, and $B_\pm = 0$. The nonzero difference $\gamma_2' - \gamma_3'$ takes into account
the cubic anisotropy of the unit cell whereas the nonzero parameters $\varkappa_0$ and $B_\pm$ reflect bulk inversion asymmetry. The $B_-$ parameter couples the functions with the opposite spin projections and, hence, is expected to be smaller than $B_+$. We neglect this parameter in the following calculations.

The parameters of the effective $6$-band Hamiltonian $\mathcal H_{\rm kp}$ can be expressed via the coupling parameters and the energy gaps  
in multi-band $\bm k$$\cdot$$\bm p$ theory. The results of such calculations in the 14-band $\bm k$$\cdot$$\bm p$ 
model~\cite{Pikus1988, Jancu2005, Durnev2014}, which includes the $\Gamma_7$ valence band and the remote $\Gamma_8'$ and $\Gamma_7'$ conduction bands in addition to the considered $\Gamma_6$ and $\Gamma_8$ bands, are summarized in Tab.~\ref{tab1}.
\begin{table*}[!ht]
\begin{center}
\begin{tabular}{c c c c c c c}
\hline
\hline
$m_0/m_e'$ &  $\gamma_1'$ & $\gamma_2'$ & $\gamma_3'$ & $\varkappa_0$ & $B_+$ & $B_-$\\
\hline  
\\
$\dfrac{2m_0 P^2}{3\hbar^2 (E_g+\Delta)}$ & $\dfrac{4m_0 Q^2}{3\hbar^2 (E_g+E_g')}$  & $-\dfrac{m_0 Q^2}{3\hbar^2 (E_g+E_g')}$ & $\dfrac{m_0 Q^2}{3\hbar^2 (E_g+E_g')}$ &
0 & $\dfrac{Q P'(E_g + 2 E_g')}{E_g'(E_g + E_g')}$ & $-\dfrac{Q P' \Delta' (E_g^2 + 2 E_g E_g' + 2E_g'^2)}{2 E_g'^2 (E_g + E_g')^2}$
\\
\hline
\hline
\end{tabular}
\end{center}
\caption{Parameters of the 6-band Hamiltonian $\mathcal H_{\rm kp}$ calculated in the 14-band $\bm k$$\cdot$$\bm p$ 
model, which includes the $\Gamma_7$ and $\Gamma_8$ valence bands, the $\Gamma_6$ conduction band, and the $\Gamma_7'$ 
and $\Gamma_8'$ remote conduction bands~\cite{Pikus1988, Jancu2005, Durnev2014}. Here, $E_g$ and $E_g'$ are the gaps at $\bm k = 0$
between $\Gamma_6$ and $\Gamma_8$ and between $\Gamma_7'$ and $\Gamma_6$, respectively, 
$\Delta$ and $\Delta'$ are the spin-orbit splittings of the valence band and the remote conduction band, 
$P$, $P'$, and $Q$ are the $\Gamma_{7,8} - \Gamma_6$, $\Gamma'_{7,8} - \Gamma_6$, and $\Gamma_{7,8} - \Gamma'_{7,8}$ coupling 
parameters, respectively. We assume that $P' \ll P$, $\Delta' \ll E_g'$ and also neglect $\bm k$-independent spin-orbit coupling 
between the $\Gamma_{7,8}$, and $\Gamma'_{7,8}$ bands.}
\label{tab1}
\end{table*}

Layers in epitaxial HgTe/CdHgTe structures are typically strained because of considerable mismatch (of about 0.3$\,\%$) between HgTe and CdTe lattice constants. The 6-band strain Hamiltonian $\mathcal H_{\rm def}$ can be constructed in a way similar to the $\bm k$$\cdot$$\bm p$ Hamiltonian~\cite{Pikus1988,BirPikus}.
Such a procedure yields the diagonal blocks
\begin{equation}\label{H66def} 
\mathcal H_{66}^{\rm def} = \Xi_c {\rm Tr} \epsilon \:,
\end{equation}
\begin{multline}\label{H88def}
\mathcal H_{88}^{\rm def} = \left(a+ \frac54 b \right) {\rm Tr} \epsilon - b \sum_{i,j} \{ J_i J_j \}_s \epsilon_{ij}  \\
+ \left( b - \frac{d}{\sqrt{3}} \right) \sum_{i \neq j} \{ J_i J_j \}_s \epsilon_{ij} \:,
\end{multline}
where $\Xi_c$ is the $\Gamma_6$-band deformation potential, $a$, $b$, and $d$ are the $\Gamma_8$-band deformation potentials,
$\mathcal H_{88}^{\rm def}$ is the Bir-Pikus Hamiltonian, $\epsilon$ is the strain tensor, and the off-diagonal blocks
\begin{equation}\label{H68def} 
\mathcal H_{68}^{{\rm def} \dag} = \left(
\begin{array}{cc}
\hspace{-9mm} - \Xi_{cv} \dfrac{\epsilon_{y'z'} -\i \epsilon_{x'z'}}{\sqrt{2}}  & \hspace{-7mm} \i \Xi_{cv}' \dfrac{2 \epsilon_{z'z'} - \epsilon_{x'x'} - \epsilon_{y'y'} }{3 \sqrt{2}}  \\
 & \\
\sqrt{\dfrac23} \, \Xi_{cv} \epsilon_{x'y'}  & - \Xi_{cv} \dfrac{\epsilon_{y'z'} -\i \epsilon_{x'z'}}{\sqrt{6}} \\
+ \i \Xi_{cv}' \dfrac{\epsilon_{x'x'} - \epsilon_{y'y'} }{\sqrt{6}}  &  \\
 & \\
\Xi_{cv} \dfrac{\epsilon_{y'z'} + \i \epsilon_{x'z'}}{\sqrt{6}}  & 
\sqrt{\dfrac23} \, \Xi_{cv} \epsilon_{x'y'}  \\
 & - \i \Xi_{cv}' \dfrac{\epsilon_{x'x'} - \epsilon_{y'y'} }{\sqrt{6}}  \\
  & \\
  \i \Xi_{cv}' \dfrac{\epsilon_{x'x'} + \epsilon_{y'y'} - 2 \epsilon_{z'z'} }{3 \sqrt{2}}   & \Xi_{cv} \dfrac{\epsilon_{y'z'} + \i \epsilon_{x'z'}}{\sqrt{2}}
\end{array}
\right) ,
\end{equation}
where $\Xi_{cv}$ and $\Xi_{cv}'$ are the inter-band deformation potentials. Note that $\Xi_{cv}$ and $\Xi_{cv}'$ vanish in centrosymmetric crystals.
The values of $\Xi_{cv}$ and $\Xi_{cv}'$ for HgTe and CdTe are not known. We expect $\Xi_{cv}'$ to be much smaller than $\Xi_{cv}$ and neglect it below.
 
Interfaces in heterostructures introduce additional mechanics of Bloch state coupling. In zinc-blende structures, interface inversion asymmetry related to the anisotropy of chemical bonds leads to light-hole--heavy-hole mixing~\cite{Aleiner1992, Ivchenko1996}.
For an arbitrary crystallographic orientation of the interface, the mixing can be described by the Hamiltonian
\begin{equation}\label{H88int}
\mathcal H_{88}^{\rm int} = \frac{\hbar^2 t_{l-h} }{\sqrt{3} a_0 m_0} \delta(\bm r \cdot \bm n + r_{\rm int}) \sum_{i} \{J_{i}J_{i+1}\}_s n_{i+2}  \:,
\end{equation} 
where $t_{l-h}$ is a dimensionless mixing parameter, $a_0$ is the lattice constant, $\bm n = (n_{x'}, n_{y'}, n_{z'})$ is the unit vector directed 
along the interface normal, say from CdHgTe to HgTe, $\bm r \cdot \bm n + r_{\rm int} = 0$ is the equation of the interface plane, and $r_{\rm int}$ is the distance between the interface and the coordinate origin. 

To describe the electron and hole states in $(0lh)$-oriented QWs we rotate the Hamiltonian~\eqref{H6band} in the reference frame 
relevant to the QW.  The transition from the reference frame $(x',y',z')$ to $(x,y,z)$ corresponds to the rotation around the $x$ axis 
by the angle $\theta$. Under this rotation, the basis functions $\ket{\Gamma_6, m}$ and $\ket{\Gamma_8, m}$ transform as the basis 
functions of the $1/2$  and $3/2$ angular momentum, respectively. Therefore, the 6-band Hamiltonian~\eqref{H6band} in the QW reference 
frame assumes the form 
\begin{equation}\label{H_xyz}
 \mathcal{H}_{\rm xyz} = R^{-1} \mathcal{H} R \,,
\end{equation}
where 
\begin{equation}
R=\begin{pmatrix}
R_6 & 0\\
0 & R_8
\end{pmatrix} \,,
\end{equation}
$R_6=\exp(i s_x \theta)$ and $R_8=\exp(i J_x \theta)$ are the $2 \times 2$ and $4 \times 4$ rotation matrices, respectively, and $s_x = \sigma_x / 2$.
The wave vector and strain tensor components are transformed as 
\begin{align}
k_{x'} &= k_x \:, \nonumber\\
k_{y'} &= k_y \cos \theta + k_z \sin \theta  \:, \nonumber\\
k_{z'} &= k_z \cos \theta  - k_y  \sin \theta \:,
\end{align}
and
\begin{align}
\epsilon_{x'x'} &= \epsilon_{xx} \:, \nonumber \\
\epsilon_{x'y'} &= \epsilon_{xy} \cos \theta + \epsilon_{xz} \sin \theta  \:, \nonumber\\
\epsilon_{x'z'} &= \epsilon_{xz} \cos \theta  - \epsilon_{xy} \sin \theta  \:, \nonumber\\
\epsilon_{y'y'} &= \epsilon_{yy}  \cos^2 \theta   + \epsilon_{zz}  \sin^2 \theta  + \epsilon_{yz} \sin 2\theta \:, \nonumber\\
\epsilon_{y'z'} &= \epsilon_{yz}  \cos 2 \theta  + (1/2)(\epsilon_{zz} - \epsilon_{yy})  \sin 2\theta \:,\nonumber\\
\epsilon_{z'z'} &= \epsilon_{zz}  \cos^2 \theta   + \epsilon_{yy}  \sin^2 \theta  - \epsilon_{yz} \sin 2\theta \:.
\end{align}

We assume that the strain in the QW is caused by mismatch between the lattice constant $a$ of the buffer layer and 
the lattice constant $a_0$ of the unstrained QW material. Then, the in-plane strain in the QW is determined from the 
lattice matching condition and given by $\epsilon_{xx} =\epsilon_{yy} =a /a_0 - 1$.
The other strain components are found from the elastic energy minimization and given by~\cite{Dantscher2015}
\begin{widetext}
\begin{align}\label{strain_in_QW}
\epsilon_{zz}&= \dfrac{
c_{11}^2 +2 c_{11} (c_{12}-c_{44})+c_{12} (-3c_{12}+10 c_{44})-(c_{11}+3 c_{12}) (c_{11}-c_{12}-2 c_{44}) \cos 4\theta
}{
-c_{11}^2-6c_{11}c_{44}+c_{12}(c_{12}+2c_{44})+(c_{11}+c_{12})(c_{11}-c_{12}-2c_{44})\cos 4 \theta
}\epsilon_{xx}\:,\nonumber\\
\epsilon_{yz} &= \dfrac{(c_{11}+2 c_{12})(c_{11}-c_{12}-2 c_{44})\sin 4\theta}{-c_{11}^2-6c_{11} c_{44}+c_{12} (c_{12}+2 c_{44})+(c_{11}+c_{12}) (c_{11}-c_{12}-2 c_{44})\cos 4 \theta}
\epsilon_{xx} \:, \;\;\; \epsilon_{xz} =0 \:,\;\;\; \epsilon_{xy} = 0\:,
\end{align}
\end{widetext}
where $c_{11}$, $c_{12}$, and $c_{44}$ are the elastic constants. Note that $c_{11}-c_{12}-2 c_{44} = 0$ in the model of isotropic elastic medium and, hence, $\epsilon_{yz} \neq 0$ due to cubic shape of the unit cell.

\begin{figure}[h!] 
\includegraphics[width=1.0\linewidth]{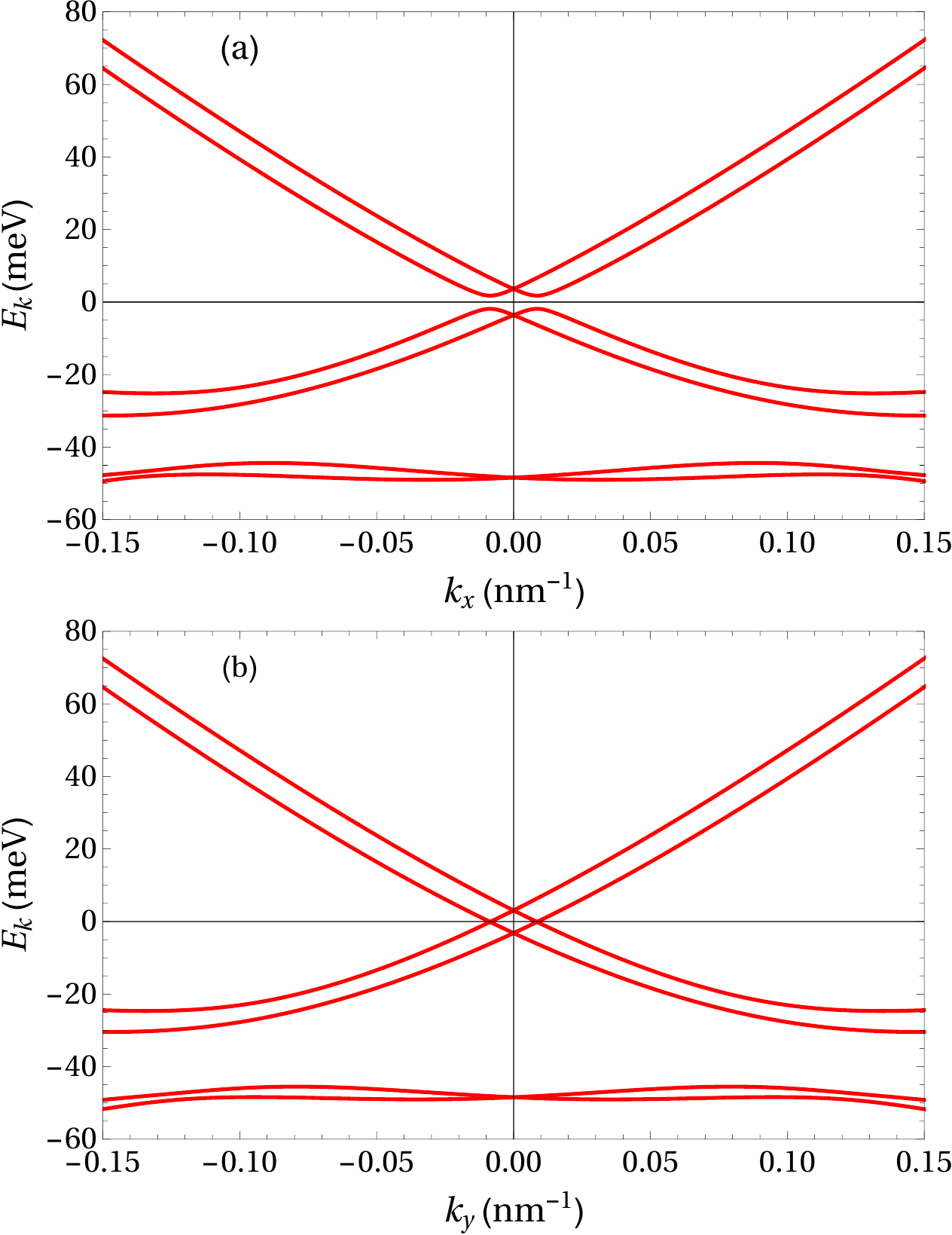}
\caption{\label{num_plot} 
Dispersions in (013)-grown HgTe/Cd$_{0.7}$Hg$_{0.3}$Te QW structure with the QW thickness $w = 6.7$~nm for the in-plane directions
(a) $\bm k \parallel x \parallel [100] $ and (b) $\bm k \parallel y \parallel [03\bar{1}] $. The dispersions are  calculated numerically in the 6-band  
$\bm k$$\cdot$$\bm p$ 
theory using HgTe and CdTe parameters from Refs.~\onlinecite{Novik2005,Dantscher2015,Tarasenko2015} and the electric field $E_z=15$~kV/cm.}
\end{figure}

Figure~\ref{num_plot} shows cross sections of the energy spectrum calculated numerically for asymmetric (013)-grown HgTe/Cd$_{0.7}$Hg$_{0.3}$Te 
QW of critical thickness. Splitting of Dirac states at $\bm k = 0$ and the in-plane anisotropy of the energy spectrum are readily seen. 

To obtain the parameters of the effective Hamiltonian~\eqref{Hqw} we solve the Schr\"{o}dinger equation 
$ \mathcal{H}_{\rm xyz}^{(\rm iso)} \Psi = E \Psi$ for zero in-plane wave vector, where $\mathcal{H}_{\rm xyz}^{(\rm iso)}$ is the isotropic part of the Hamiltonian~\eqref{H_xyz}, and find the functions $\ket{E1, \pm 1/2}$ and $\ket{H1, \pm 3/2}$. Then, we project the Hamiltonian 
$( \mathcal{H}_{\rm xyz} - \mathcal{H}_{\rm xyz}^{(\rm iso)})$ onto the basis states $\ket{E1, \pm 1/2}$ and $\ket{H1, \pm 3/2}$. 
This procedure yields the effective Hamiltonian~\eqref{Hqw} with $A = P/\sqrt{2} \int f_1(z) f_3(z) dz$, 
\begin{eqnarray}
\label{splitting_ans}
\eta &=&  \frac{\hbar^2 t_{l-h}}{2m_0 a_0} \left[ f_3(w/2) f_4(w/2) - f_3(-w/2) f_4(-w/2) \right]  \\
&+&  \frac{1}{\sqrt{2}} \int dz f_1(z) \Xi_{cv} \left[\epsilon_{yz} \cot 2\theta + \frac{\epsilon_{zz} - \epsilon_{yy}}{2}\right] f_3(z) \nonumber \\
&-&\frac{1}{2 \sqrt{2}} \int dz f_1(z) \partial_z  B_+ \partial_z f_3(z) \:, \nonumber \\ \nonumber \\
\gamma &=& \frac{\hbar^2 t_{l-h}}{2m_0 a_0} \left[ f_3(w/2) f_4(w/2) - f_3(-w/2) f_4(-w/2) \right] \:, \nonumber \\ \nonumber \\
\chi &=& \frac{\sqrt{3} \hbar^2}{4m_0} \int dz f_4(z) \partial_z (\gamma_2' - \gamma_3') \partial_z f_3(z) \nonumber \\
&-&\frac12 \int dz f_3(z) \left\{ \frac{ \left[(d + \sqrt{3} b)  + (d - \sqrt{3}b) \cos 4\theta \right] \epsilon_{yz}}{\sin 4 \theta}  \right. \nonumber \\
&&\left. + (d - \sqrt{3}b) \frac{\epsilon_{zz} - \epsilon_{yy}}{2} \right\} f_4(z) \:, \nonumber \\ \nonumber \\
\zeta &=& \frac{\sqrt{3} \hbar^2}{4m_0} \int dz f_4(z) \partial_z (\gamma_2' - \gamma_3') \partial_z f_3(z) \nonumber \\
&+& \frac{1}{2} \int dz f_3(z) (\sqrt{3} b - d) \left[\epsilon_{yz} \cot 2\theta + \frac{\epsilon_{zz} - \epsilon_{yy}}{2}\right] f_4(z) \nonumber \:,
\end{eqnarray}
where $w$ is the QW thickness and the strain tensor $\epsilon$ is given by Eq.~\eqref{strain_in_QW}. The 
 strain-induced contributions to $\eta$, $\chi$, and $\zeta$ depend on the $\theta$ angle. 
 Nevertheless, the major effect of QW crystallographic orientation on the fine structure of Dirac states 
is captured by Eqs.~\eqref{HIIABIA} and~\eqref{HSIA}. Note that, in QWs with symmetric confinement, $\chi = \zeta = 0$ since $f_3(z)$ and $f_4(z)$ have opposite parities.

Finally, we estimate the coupling parameters $\gamma$, $\eta$, $\zeta$, and $\chi$ for HgTe/Cd$_{0.7}$Hg$_{0.3}$Te QWs with the thickness $w = 6.7$~nm using the band parameters, elastic constants, and interface mixing from Refs.~\onlinecite{Novik2005,Dantscher2015,Tarasenko2015}.  
The estimation shows that the dominant contribution to $\gamma$ and $\eta$
of approximately $5$~meV comes from IIA with $t_{l-h} \sim 1$. The strain contribution to $\eta$ is of the order of 1 meV for the interband deformation potential $\Xi_{cv} \approx -1$~eV and the strain tensor components $\epsilon_{yy} = 3\cdot 10^{-3}$, $\epsilon_{zz} = -4\cdot 10^{-3}$ and $\epsilon_{yz}=10^{-3}$ calculated after Eq.~\eqref{strain_in_QW}. The BIA contribution to $\eta$ is $\sim 0.2$~meV for the parameter $B_+ \approx 0.4 \hbar^2/m_0$ estimated following Tab.~\ref{tab1} for $E_g' = 4.5$~eV~\cite{Lu1989}, $Q = P$, and $P' = 0.1~P$. The $\chi$ and $\zeta$ parameters are related to SIA and estimated for the electric field $E_z=15$~kV/cm$^2$. The contribution to $\chi$ and $\zeta$ from the cubic warping of the energy spectrum determined by $\gamma_2' - \gamma_3'$ is of the order of $0.1$~meV. The strain contributions to $\chi$ and $\zeta$ are about $-0.1$~meV and $-0.2$~meV, respectively. Note that SIA-related coupling parameters scale with the electric field $E_z$ and can be larger.

\section{Summary}

To summarize, we have studied theoretically the fine structure of Dirac states in HgTe/CdHgTe QWs of critical and close-to-critical thicknesses.
Taking into account bulk, interface, and structure inversion asymmetry in these zinc-blende-type QWs, we have derived the effective Hamiltonian that describes the splitting of Dirac states in the class of $(0lh)$-oriented QWs from a unified standpoint. The Hamiltonian contains four parameters of splitting at zero in-plane wave vector. We have calculated these parameters in the 6-band extended Kane model, which takes into account the lack of inversion center
and the cubic shape of the host crystal lattice, the elastic strain in the QW, and the heavy-hole--light-hole mixing at the QW interfaces. 
Further, we have derived an analytical expression for the energy spectrum of the Dirac states as a function of the growth direction 
and studied how the spectrum evolves from (001)- to (013)- and (011)-grown QWs. In general case, the spectrum is anisotropic and in the QWs of critical thickness contains four Weyl points. The positions of the Weyl points depend on the QW crystallographic orientation and structure inversion asymmetry
and can be controlled by an external electric field applied along the QW normal, e.g., by gate voltage.

\acknowledgements 

This work was supported by the Russian Science Foundation (project 22-12-00211).
G.V.B. acknowledges the support from the ``BASIS'' foundation.

\end{document}